\documentclass{article}

\usepackage[english]{babel}
\usepackage[utf8]{inputenc}
\usepackage[letterpaper,top=2cm,bottom=2cm,left=3cm,right=3cm,marginparwidth=1.75cm]{geometry}

\usepackage{amsmath}
\usepackage{graphicx}
\usepackage{natbib}
\usepackage{csquotes}
\usepackage{fixltx2e}
\usepackage{gensymb}
\usepackage{ccaption}

\usepackage{hyperref}

\title{Predictive Geological Mapping with Convolution Neural Network Using Statistical Data Augmentation on a 3D Model}
\author{Cedou Matthieu, Gloaguen Erwan, Blouin Martin, Caté Antoine,\\ Paiement Jean-Philippe, Tirdad Shiva}

\begin{document}
\maketitle

\begin{abstract}

Airborne magnetic data are commonly processed and interpreted to produce preliminary geological maps. Machine learning has the potential to partly fulfill this task rapidly and objectively, as geological mapping is comparable to a semantic segmentation problem that can be solved with a convolutional neural network. Because this method requires a high-quality dataset, we developed a data augmentation workflow that uses a 3D geological and petrophysical (magnetic susceptibility) model as input. The workflow uses soft-constrained Multi-Point Statistics, to create many synthetic 3D geological models, and Sequential Gaussian Simulation algorithms, to populate the models with the appropriate magnetic distribution. Then, forward modeling is used to compute the airborne magnetic responses of the synthetic models, which are associated with their counterpart surficial lithologies. We applied this workflow on a 3D model of the geology and magnetic susceptibility of the Malartic Mine area to obtain a large airborne magnetic synthetic dataset, associated with surficial lithologies, and perform segmentation. A Gated Shape Convolutional Neural Network algorithm was trained on this synthetic dataset to perform geological mapping of the synthetic airborne magnetic data and detect lithological contacts. The algorithm also provides attention maps highlighting the structures at different scales, and clustering was applied to its high-level features to do a semi-supervised segmentation of the area. The validation conducted on a portion of the synthetic dataset shows that the methodology is suitable to segment the surficial geology using airborne magnetic data. We also used transfer learning to test the trained model on measured magnetic data from adjacent areas, without re-training. The clustering shows a good segmentation of the magnetic anomalies into a pertinent geological map in these areas. Moreover, the first attention map isolates the structures at low scales and shows a pertinent representation of the original data. The quality of the results empirically validates our data augmentation method. Furthermore, it proves that using convolutional neural networks is pertinent for preliminary geological mapping. Thus, our method can be used in any area where a geological and petrophysical 3D model exists to train a deep learning algorithm and produce preliminary geological maps of good quality and new representations of the input data in any area sharing the same geological context, using airborne magnetic data. (github: \url{https://github.com/MatthieuCed/GSCNN-apply-to-airborne-magnetic})

\end{abstract}

\section{Introduction}

Geological maps are crucial as a means to gather all the available information of an area in order to understand its geological setting. The maps represent an interpretation of the surficial geology and summarize the understanding of a geological terrain in a 2D image. To produce them, geologists use field observations, geological knowledge of the regional geology, and they often use airborne geophysical data to refine their understanding of the zone of interest. \citep{soller_geologic_2004, jaques_high-resolution_1997}.

\medskip

Indeed, high-resolution airborne geophysical methods, such as airborne magnetic, are of great use in geological mapping as they provide fast, cost-effective, and spatially continuous data over large and remote areas. They permit to measure the spatial variations of physical properties of a terrain, such as magnetic susceptibility, detect discontinuities in those properties, thus provide indirect but continuous information of the underlying geology \citep{murthy_airborne_2007}. For example, geologists use magnetic data to determine basement structures and continuities of geological terrains \citep{jaques_high-resolution_1997}.

\medskip 

Because some remote areas do not have enough field observations available to produce traditional preliminary cartography, some authors in the literature propose using airborne geophysical data and machine learning algorithms to produce such maps. This process is known as predictive geological mapping in Earth Sceinces but is considered as a semantic segmentation in Computer Vision. For example, \cite{harvey_geological_2016} compared Naïve-Bayes, K-Nearest Neighbour, Random-Forest, and Support Vector Machine algorithms to generate preliminary geological maps. The Geological Survey of Canada also used Random Forest, Maximum Likelihood Classification and Artificial Neural Networks to predict geology in Canada's Northern areas (\citep{harris_predictive_2015, schetselaar_remote_2007}). \cite{hood_improved_2019} used Random-Forest on Principal Component Analysis' representations from different airborne geophysical data and features calculated from those data to predict lithologies. \cite{carter-mcauslan_predictive_2021} also used Self Organizing Maps to do the predictive mapping of an area with magnetic, radiometric, and airborne gravity data. 

\medskip

The strengths of the machine learning approach are rapidity and objectivity. It provides a first preliminary map that could drive early decision-making and can be reproduced and standardized, unlike handmade cartography \citep{cracknell_geological_2014}. 

\medskip

All of these studies, used shallow learning algorithms that require a large number of discriminative and handcraft features as input data to perform \citep{chollet_deep_2018}. Thus, those authors had to generate their own features from the limited available data, and are restricted to use these algorithms on the areas where different airborne geophysical datasets are available, such as magnetic, radiometry, and gravity \citep{cracknell_geological_2014}.

\medskip

However, supervised Deep Learning approaches took over in the segmentation field of research over the last decade, with algorithms such as Convolutional Neural Networks (CNN), Encoder-Decoders Models, Recurrent Neural Networks, Generative Adversarial Networks \citep{minaee_image_2020}. Comparatively to the shallow algorithms, deep learning segmentation algorithms applied to computer vision tasks generate many representations from an input image and isolating their main features \citep{chollet_deep_2018}. Those representations account for the spatial layout of the data at different scales and are adapted to the input images during the training. In parallel, these deep learning algorithms use those representations to make their predictions when new images are presented to them \citep{minaee_image_2020}. However, deep learning segmentation algorithms have to be trained on a dataset containing a large variety of possible scenarios. Thus, training deep learning algorithms requires a large amount of training data, with high-quality hand-crafted training labels \citep{vanderplas_python_2016}. Indeed, the quality of the output of such an algorithm depends on the quality of its training dataset \citep{chollet_deep_2018}.

\medskip

The interpretive nature of geological maps, their heterogeneous resolutions, and the sparsity of the outcrop where the geology is validated limit the use of existing geological maps to train deep learning algorithms.  The validation of the results is also tricky because of the same reasons. Because of this label-limitation problem, most of the approaches in the literature \citep{carter-mcauslan_predictive_2021, harris_predictive_2015, schetselaar_remote_2007} use shallow unsupervised clustering algorithm, which finds the natural classes in the data, and where no label is required.

\medskip

To answer the label-limitation problem, we propose in this paper a data augmentation workflow using a known 3D conceptual geological and petrophysical measurements to train a segmentation algorithm. From this conceptual geological model, many equiprobable synthetic geological models are generated using the Multi-Point Statistics algorithm. Then, those models are filled with the appropriate magnetic susceptibility data distribution using Sequential Gaussian Simulations. At last, the airborne magnetic responses of each model are calculated by forward modeling. Once generated, those responses, associated with their counterpart surficial geological maps, can be used to train a deep learning algorithm. The 3D model used to test our approach is from the extensively explored area of the Malartic Mine (see figure \ref{fig:localisation}), located in the Abitibi subprovince of Quebec, Canada. Using transfer learning hypothesis, the methodology has been applied to real data of the close to the same area. 

\section{Study Area} 

Applying deep learning to predictive geological mapping using airborne magnetic data requires training areas with a similar geological setting and precise cartography, to be used as training labels. However, geological maps are produced by an interpretive process, are subject to change with new acquired knowledge or data, and are at different resolutions than the magnetic data. Thus, only high-quality geological maps can be used to train a deep learning algorithm with confidence which rarely really exists in real life as outcrops are often in limited numbers.

\medskip

To get around this problem, we propose a geostatistical data augmentation workflow, to generate many training examples pairs of geological surficial maps and airborne magnetic data. 

\medskip

In order to test the data augmentation workflow, we used a 3D lithological model and its magnetic susceptibility dataset of an extensively explored area in the province of Québec, Canada, namely the Malartic area. The geological context of the model's area and the model itself are presented in details in the next sections.

\subsection{Geological Context Of The Malartic Area} \label{chap:geo}

\begin{figure}
\centering
\includegraphics[width=1\textwidth]{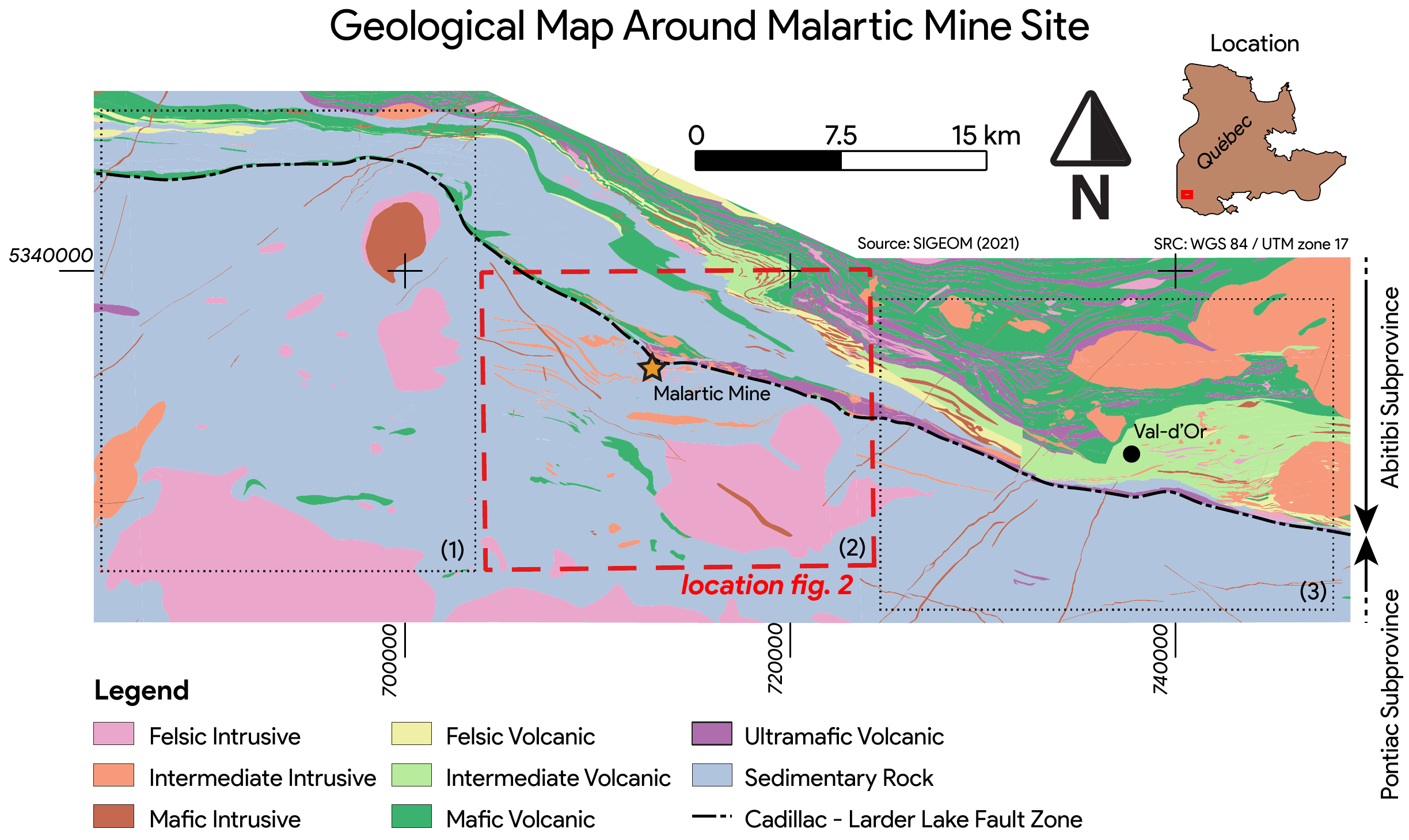}
\caption{\label{fig:localisation}Simplified geological map around the Malartic mine site following the Cadillac-Larder Lake Fault (black dash-dotted line), location of the 3D model of the Malartic mine area (red dash-dotted line), location of the testing data (thin black dotted line), and location of the area in Québec province \citep{sigeom_sigeom_2019}}
\end{figure}

The study area is located in the southern part of the Abitibi subprovince, in Québec province, Canada. It covers a 20 km by 14 km area, oriented East-West. It is centered on the Malartic open-pit Mine. Figure \ref{fig:localisation} shows the location of the area, the Malartic Mine, and the surrounding geology.

\medskip

The area is located in the Archean Superior province of Canada. It lies astride the Abitibi subprovince (to the North) and the Pontiac subprovince (to the South). The Cadillac-Larder Lake Fault Zone, a large trans-crustal fault associated with numerous orogenic-gold deposits, separates the two provinces (CLLFZ, black dash-dotted line on figure \ref{fig:localisation}) \citep{rabeau_gold_2010}.

\medskip

The Abitibi subprovince is located north of the CLLFZ (see figure \ref{fig:localisation}). It is characterized by volcanic episodes dated between 2765 to 2695 Ma and displays a low metamorphic grade (greenschist facies) \citep{thurston_depositional_2008}. The study area shows, from North to South (see figure \ref{fig:original_model}.1): the Malartic group, composed of volcanic assemblages and plutonic ultramafic rocks and having a strong magnetic signature; the Kewagama group, composed of greywackes; the Hébecourt formation, composed of basalt andesites and gabbro; the Cadillac group, composed of greywackes and conglomerates; and the Piché group, in the CLLFZ, composed of volcanic and ultramafic rocks having a high magnetic response \citep{de_souza_geology_2015, de_souza_geology_2017}. As shown in figure \ref{fig:original_model}.2, Kewagama, Hébécourt, and Cadillac groups show weak magnetic responses.

\medskip

The Pontiac subprovince is in the southern part of the study zone, South of the CLLFZ. It is dominated by a greywackes sequence, similar to the Kewagama group, dated between 2685 to 2682 Ma, and displays a high metamorphic grade from greenschist to amphibolite facies \citep{de_souza_geology_2015, camire_archaean_1993}. The Pontiac also includes mafic and ultramafic metavolcanic rocks \citep{camire_archaean_1993}, and granodiorite to diorite intrusions \citep{de_souza_geology_2015}. As figure \ref{fig:original_model} shows, some late Proterozoic diabase dykes (mafic intrusive in figure \ref{fig:localisation}) cut the studied area in a northwest-southeast direction \citep{buchan_diabase_2004}.

\medskip

The Malartic mine is a gold deposit located in the metasediments of the Pontiac group. At the mine scale, dykes and sills of various compositions (diorite-monzodiorite in figure \ref{fig:original_model}) cut the sediments. The mineralization fluids were channeled by the CLLFZ and trapped in the sediments in the form of quartz-pyrite stock-work or disseminated pyrite \citep{de_souza_geology_2015}. 

\subsection{Malartic Lithological and Petrophysical Model}

\begin{figure}
\centering
\includegraphics[width=1\textwidth]{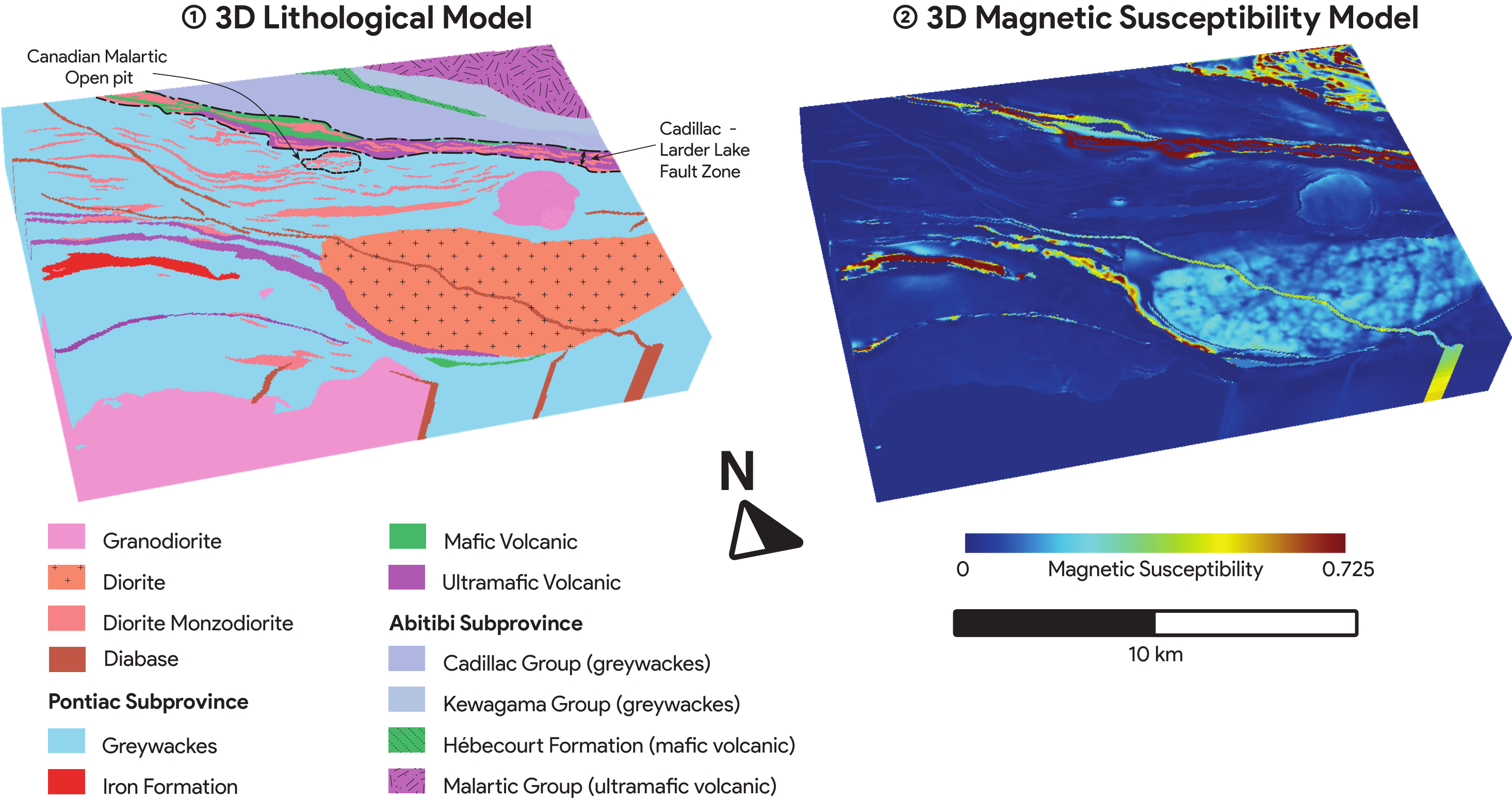}
\caption{\label{fig:original_model}(1) The 3D lithological model (281$\times$401$\times$56 cells along North, East and vertical axis; cells of 50$\times$50$\times$50 m) provides by Mira Geoscience Ltd. on Malartic mine site with Cadillac-Larder Lake fault zone and Malartic open pit mine delimited; (2) same 3D model populated with magnetic susceptibility provided by Mira Geoscience Ltd. \citep{vallee_geophysical_2019}}
\end{figure}

The 3D geological model used in this paper is a preexisting geological model of the Malartic Mine area. The technical information about the model can be found in \cite{vallee_geophysical_2019}. The location of the model is shown in Figure \ref{fig:localisation}.2. The model is built from a 3D grid (voxet) containing lithological and magnetic susceptibility values at each cell (voxel, figure \ref{fig:original_model}). The size of the voxels is $50 \times 50 \times 50$ m. The grid contains 401 cells along the East-West direction, 281 cells along the South-North direction, and 56 cells in the vertical direction, for a total of 6 310 136 cells \citep{vallee_geophysical_2019}. 

\medskip

The lithological model (figure \ref{fig:original_model}.1) comprises 12 lithologies (described in section \ref{chap:geo}.1). It was generated using the known cartography of the area, the pit mapping data of the Malartic Mine, the core-logging description, and the geophysical data available on the area \citep{vallee_geophysical_2019}. 

\medskip

The petrophysical model (figure \ref{fig:original_model}.2) was obtained by constraint 3D inversion. A starting magnetic susceptibility value was assigned to each model cell using known geological contacts, borehole measurements, geophysical interpretation, and physical proprieties analysis. Then, the model was refined for both magnetic and geological values: (1) During the inversion, each cell's magnetic property is changed until the difference between the calculated magnetic response of the model and the known response is optimum. (2) Once the inverted model is obtained, the user changed the geological model to fit the known geological setting. Then, the new geological model is used again to constraint the magnetic inversion, and the process iterates until the optimum geological model and magnetic response are found \citep{fullagar_towards_2007, vallee_geophysical_2019}. 

\section{Methodology}

\subsection{Geostatistical Data Augmentation}

Our Geotatistical Data Augmentation module aims generating realistic synthetic airborne magnetic maps and their corresponding surficial geology in order to built a dataset enabling the training of a deep learning architecture. It is composed of 3 modules described below. The implementation details are described in part \ref{chap:imp}.

\subsubsection{Multi-Point Statistics}

\begin{figure}
\centering
\includegraphics[width=1\textwidth]{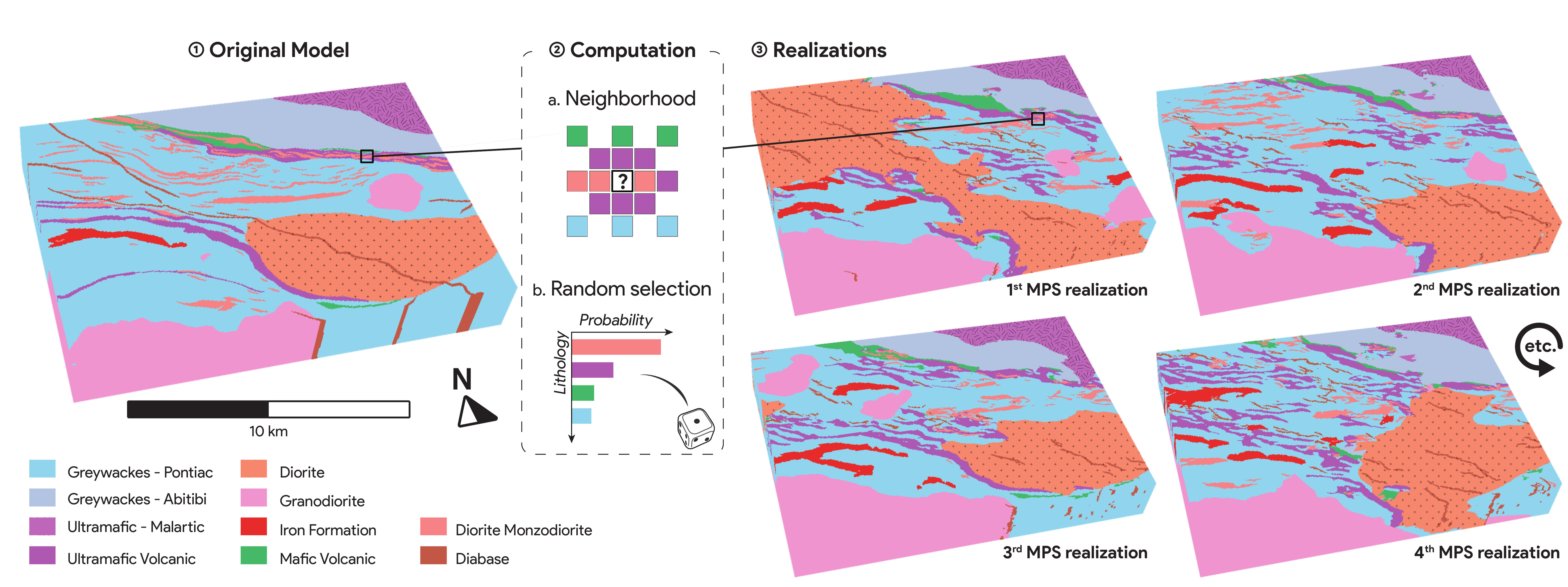}
\caption{\label{fig:MPS}Illustration of \textit{Multi-Point Statistics} (MPS) method: from an original model (1), the MPS extracts patterns (or neighborhood) (2.a) and compute the probability of the nature of the central point, according to each scenario (2.b) ; each point of the simulations (3) is generated by a random sampling on the computed probabilities, according to the known points of its neighborhood \citep{mariethoz_multiple-point_2014}}
\end{figure}

The Malartic 3D conceptual geological model \citep{vallee_geophysical_2019} is used as an input to compute a high number of equiprobable geological 3D models using Multi-Point Statistics algorithm (MPS) \cite{guardiano_multivariate_1993}. 

\medskip

Indeed, MPS allows for the generation of multiple stochastic categorical model, based on an existing categorical model as an input example, also called training image (\ref{fig:MPS}.1 shows the conceptual lithological model of Malartic). MPS extracts every pattern found in this 3D training image, following a multiple scale template approach (or neighborhood, illustrates in figure \ref{fig:MPS}.2.a) and saves them in a database \citep{mariethoz_multiple-point_2014}.

\medskip

To obtain a realization, MPS sequentially computes a categorical value, for each voxel of the grid to fill, following a predefined path. It interpolates lithologies between control points, where the lithologie is known. MPS considers, for each voxel to be simulated, its neighborhood composed of previously simulated voxels and control points, following the given template. The center value of each scenario, similar to the observed one, is summed to obtain a probabilistic distribution (illustrated in figure \ref{fig:MPS}.2.b). The empty voxels present in the neighborhood are considered as if they can contain every possible value. The final value is obtained by a random selection based on the computed distribution and added to the synthetic model being filled. This process is repeated until the whole grid is populated \citep{mariethoz_multiple-point_2014}. Figure \ref{fig:MPS}.3 shows four realizations obtained by MPS of the Canadian Malartic geological model. 

\subsubsection{Simulation of the physical properties}

\begin{figure}
\centering
\includegraphics[width=1\textwidth]{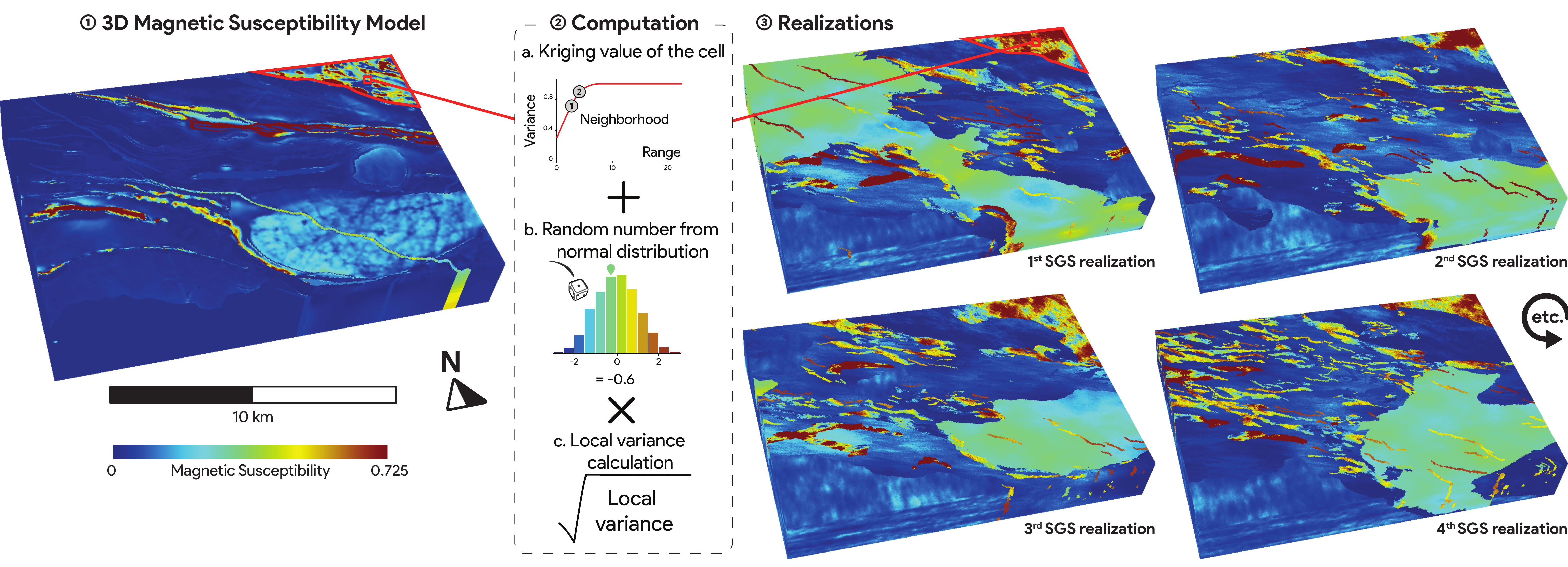}
\caption{\label{fig:SGS}Illustration of the \textit{Sequential Gaussian Simulation} (SGS) algorithm: the original petrophysical model (1) is decomposed in regions of same lithology, assuming they share the same petrophysical propriety ; the semi-variogram is defined on the values of the region of interest (2.a) ; each voxel of each geological object contains in each simulated model (3) is generated by computing the Kriging value calculated on its neighborhood (2.a) plus a random Gaussian noise (2.b) multiplied by the local variance (2.c) \citep{nowak_practice_2005}. The model populated are the synthetic geological model from figure \ref{fig:MPS}}
\end{figure}

Once multiple geological scenarios of the geology are computed using the MPS approach, one has to be populated with realistic magnetic susceptibility spatial distribution to compute their airborne magnetic response.

\medskip

Sequential Gaussian Simulation (SGS), proposed by \cite{matheron_intrinsic_1973}, is used to fulfill this task \citep{nowak_practice_2005}.

\medskip 

Figure \ref{fig:SGS} illustrates the SGS algorithm, using the Malartic petrophysical model (\ref{fig:SGS}.1) as input data. Because SGS requires stationary data (data respecting the same semi-variogram at every point of the model; \citealp{nowak_practice_2005}), each lithological domain was computed separately. During simulations, each voxel is visited sequentially, following a random path. The conditional mean and variance are computed using kriging at each voxel location (see figure \ref{fig:SGS}.3) based on its neighbors: previously-calculated voxels or previously measured values. Then, the final value of the voxel to simulate is obtained by adding a randomly picked value from the conditional distribution (2.b) \citep{nowak_practice_2005}. The four \textquote{SGS realization} examples in figure \ref{fig:SGS}.3 were simulated using this algorithm within each lithological domain of the geological models obtained by MPS shown in figure \ref{fig:MPS}.

\medskip

In order to further augment the variance of the synthetic data, the magnetic susceptibility distribution and variogram can be randomly assigned to the lithological domain to fill. This step is particularly important to generalize the training than will enable transfer learning to real data and to other similar geological environment elsewhere.

\subsubsection{Magnetic Forward Modeling}

\begin{figure}
\centering
\includegraphics[width=1\textwidth]{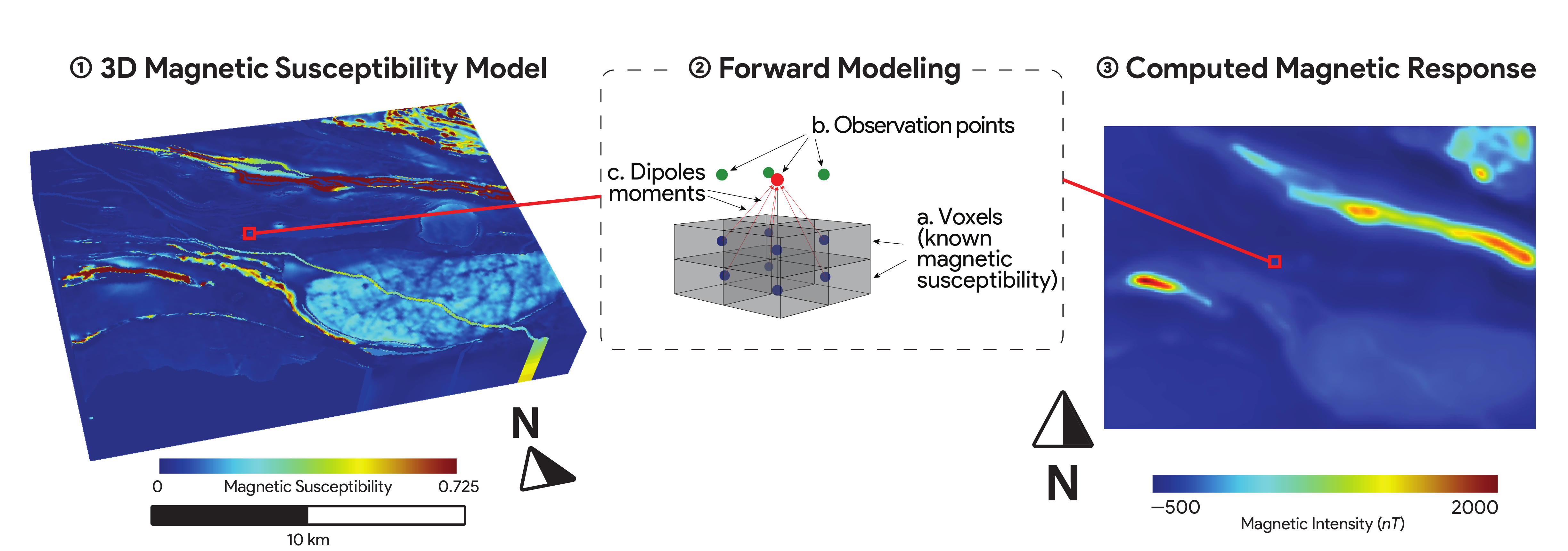}
\caption{\label{fig:modelisation}Computing the geophysical response of the original 3D magnetic model (1) by doing the sum of the dipoles moments (2.c) of each voxel (2.a) at an observation point (2.b); repeating the method at each observation point provides an airborne residual magnetic-like response (3) \citep{lelievre_magnetic_2006}}
\end{figure}

Once the geological and magnetic susceptibility 3D models are calculated, their airborne magnetic response must be computed, and their counterpart surficial geology must be extracted to get the desired training set output. 

The surficial geology maps are obtained by extracting the surficial values of the synthetic lithological models generated by MPS. 

\medskip

The magnetic forward modeling is computed on the synthetic magnetic susceptibility 3D models using the algorithm developped by \citep{lelievre_magnetic_2006}. The magnetic intensity, induced by the magnetic susceptibility values of a model (figure \ref{fig:modelisation}.1) projected in a regular 3D grid (\ref{fig:modelisation}.2.a), is computed at given observation points (red point in figure \ref{fig:modelisation}.2.b) as the sum of the magnetic moment (figure \ref{fig:modelisation}.2.c) of the grid \citep{lelievre_magnetic_2006}. 

\medskip

Once each magnetic intensity is computed for each observation point, the result is similar to a residual airborne magnetic survey. Then, the obtained synthetic airborne magnetic data are normalized, and extreme values are removed. Figure \ref{fig:modelisation}.3 shows the modeling result of the Malartic petrophysical model. 

\subsection{Segmentation With Gated Shape CNN}

A deep learning algorithm can now be trained on the synthetic airborne magnetic and surficial geology data generated with the data-augmentation workflow. 

\medskip

Generic deep learning architecture applied to segmentation problems are composed of convolutional layers, containing spatial filters, which are optimized during training by a gradient descent method, to extract features of interest (in feature maps) from an input image. Those algorithms use those feature maps to make their predictions, from multiple non-linear operations \citep{minaee_image_2020}. The output is a segmented map of the input geophysical image, containing a categorical value for each pixel. 

\medskip

The algorithm selected is the Gated-Shaped CNN (GSCNN) proposed by \cite{takikawa_gated-scnn_2019} to perform semantic segmentation on urban scenes from the CityScape dataset \citep{cordts_cityscapes_2016}. Its power lies in the fact that it simultaneously conducts contact detection, and semantic segmentation.  

\subsubsection{Architecture of Gated Shape CNN}\label{chap:gscnn}

\begin{figure}
\centering
\includegraphics[width=1\textwidth]{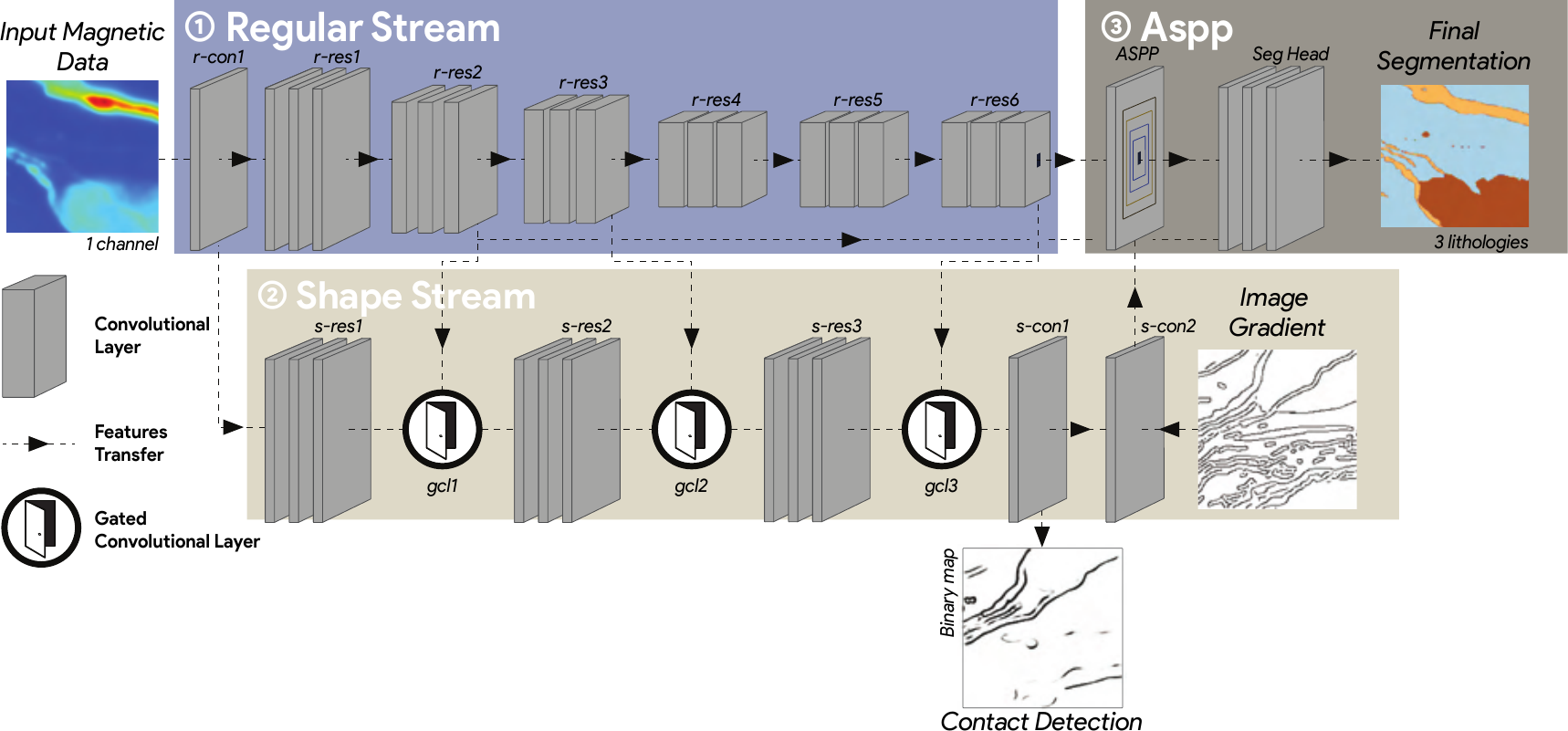}
\caption{\label{fig:gscnn} The architecture of the Gated-Shape CNN segmentation model used: a Wide-ResNet backbone (1 in figure), which contains 1 convolutional layer (\textit{r-con1}) and 6 Residual Blocks (\textit{r-res1} to \textit{r-res6}), provides deep features representation used by the Shape Stream (2), composed of 3 Residual Blocks (\textit{s-res1} to \textit{s-res3}) and 3 Gated Convolutional Layers (\textit{gcl1} to \textit{gcl3}), to detect the contacts of the input image; and by a fusion module (3) which combines the backbone and the contacts information, using an Atrous Spatial Pyramid Pooling module (ASPP), to compute the final segmentation \citep{takikawa_gated-scnn_2019}}
\end{figure}

The architecture of the GSCNN algorithm is presented in figure \ref{fig:gscnn}. The idea behind this algorithm is to channel the information into two-stream: (1) a \textquote{Regular Stream} that extracts intensity and textural information, (2) and a \textquote{Shape Stream} that extracts the contact information \citep{takikawa_gated-scnn_2019}.

\paragraph{Regular Stream}

The Regular Stream (1 in figure \ref{fig:gscnn}) is a standard segmentation CNN. It produces feature maps from an input image. In the current implementation, the algorithm used is a Wide-ResNet \citep{zagoruyko_wide_2017}. The Wide-ResNet contains \textquote{Drop-out} layers, which randomly ignore some convolutional layer outputs, in a ResNet architecture. The ResNet contains \textquote{Skip-layers} between blocks of convolutional layers (known as \textquote{Residual Blocks}). Those layers work as an identity function, allowing all information to flow across the model and let the algorithm optimizes the residual information to fulfill its task \citep{he_deep_2015}. 

\paragraph{Shape Stream}

The aim of the Shape Stream (2 in figure \ref{fig:gscnn}) is to channel the information of the objects' contacts. It uses the feature maps produced by the Regular Stream to extract the information related to contacts. It is a shallow network, operating on the input image resolution, which produces a contacts detection map to be compared with a known contacts mask \citep{takikawa_gated-scnn_2019}.

\medskip

To extract contacts-relevant-only information, \textquote{Gated Convolutional Layers} (GCL) are interleaved in the Shape Stream architecture, as shown in figure \ref{fig:gscnn}. GCL produces 2D attention maps using feature maps from the Regular Stream, highlighting the area of importance for the contacts (between 0 and 1) at different scales. Those attention maps are multiplied by the information contained in the convolutional layers of the Shape Stream. Thus, only information of interest for contacts detection passes through the Shape Stream model \citep{takikawa_gated-scnn_2019}. The output of the Shape Stream is a binary map of the presence or absence of contact for each pixel. 

\paragraph{Atrous Spatial Pyramidal Pooling}

\textquote{Atrous Spatial Pyramidal Pooling} (ASPP, 3 in figure \ref{fig:gscnn}) is a convolutional layer which aggregates features from different scales. It computes convolutions at different sampling rates defined by the user, to account for the different feature sizes observed in the data. The products of those convolutions are stacked in its output \citep{chen_deeplab_2017}. The output of the ASPP is merged with the output of the first residual block of the Regular Stream (\textit{r-res1} in figure \ref{fig:gscnn}) and sent to the last segment of the network (\textit{Seg Head} in figure \ref{fig:gscnn}) composed of 3 convolutional layers to perform the final semantic segmentation. 

\paragraph{Loss Functions}

The Regular and Shape Streams are jointly trained. The output of the Shape Stream, the predicted contacts, is compared with the borders of the ground truth segmentation using a binary cross-entropy loss. The output of the semantic segmentation head is compared with the ground truth segmentation using a cross-entropy loss. At last, a \textquote{Dual Task Loss} is introduced to ensure that contacts of the final segmentation correspond to the ground truth contacts \citep{takikawa_gated-scnn_2019}.

\paragraph{Clustering}

In the GSCNN, the last part of the network (\textit{Seg Head} in figure \ref{fig:gscnn}) classifies the pixels of an input image using the feature maps of the ASPP module. Those feature maps consider different spatial resolutions, providing an understanding of the scene at different scales \citep{chen_deeplab_2017}. 

\medskip

Those feature maps are stacked into a multi-band image (the number of bands corresponding to the stacked feature maps). Then, a clustering algorithm classifies each pixel of this image into natural groups. The product of this clustering is a semi-supervised semantic segmentation of the input image.

\medskip

The clustering algorithm used is the Hierarchical Clustering \citep{murtagh_wards_2014}, using the ward linkage method, which minimizes the variance for each cluster. Before clustering, a Principal Component Analysis (PCA, see \citealp{jolliffe_principal_2002}) is applied on the 304 feature maps of the ASPP's output to reduce computing time. It reduces the number of channels to the number of Principal Components required to explain 95\% of the ASPP's variance.

\paragraph{Lithological Domain Simplification}

Deep learning algorithms only consider the shapes, intensity, and textures of the signal they are processing. They do not have an understanding of the objects, their nature, nor their forming process \citep{chollet_deep_2018}.

\medskip

During training, the CNNs have to learn on a high variance database. If the training scenarios are too similar, the algorithm will over-fit the training data, and will not be able to perform on new data \citep{chollet_deep_2018}. To prevent this scenario, the magnetic susceptibility distributions can be mix in the SGS module of the Data-Augmentation.

\medskip

At last, if a class is under-represented in a training database, the algorithm might over-fit or ignore it \citep{chollet_deep_2018}. If possible, such class has to be removed or merge with another one. 

\medskip

We then defined the lithological domains by merging lithologies into training groups following their similarity in terms of magnetic susceptibility. Those new groups might not respect geological facies, but are magnetic facies.

\subsection{Implementations Details}\label{chap:imp}

\paragraph{Data Augmentation}

The MPS and the SGS were run in SKUA-GOCAD$^{TM}$ 2017. Because the SKUA-GOCAD$^{TM}$'s MPS algorithm implementation takes a maximum of 10 lithologies, the Cadillac group, the Kewagama group, and the Hébécourt formation were merged because of the similarity of their shape and their magnetic signature. 
\medskip 

The MPS' control points are defined by fifteen random \textquote{drillholes} passing through the model, perpendicular to the geological structure (N40E 45SW). The MPS template is defined by six sub-grids in 3D, with a maximum search ellipsoid of 12 cells and 62 neighbors. 

\medskip

The semi-variograms used in the SGS was saled to the observed distributions of the magnetic susceptibility values of the merged lithological domains. No control point was added to the SGS simulations in order to generalize the training. 

\medskip

 A first database of 195 models was created without mixing the magnetic distributions. The realizations were calculated on 39 drillholes scenarios, and 1 SGS simulation was computed on each scenario. A second database of 150 models was created by randomly mixing the variograms and distributions of the magnetic susceptibility of each lithological domain by pools of plutons, dykes, and greywackes. 75 MPS realizations were computed, with 15 different drillholes scenarios, and 2 SGS simulations were run on each MPS realization. 
 
 \medskip
The forward airborne magnetic modeling was conducted with Mira Geosciences ltd.'s MAGFOR3D algorithm. This step is particularly computer intensive as the forward modeling took around 3:40 h for each modeling pass on a 2 $\times$ Intel Xeon (2.67GHz) CPU machine with 48 GB RAM.

The observation points of the forward modeling are 140 meters above the model's surface and spaced by 50 meters in East-West and North-South directions to mimic real airborne magnetic surveys.

\paragraph{GSCNN} 

The model is implemented in Pytorch and was trained under Google Colab$^{TM}$, using GPU and high memory RAM (1 Tesla V100 SXM2 16 GB). The learning rate was set at 0.002 and decreased by 0.0001 per epoch during training. A coefficient of 1 was applied to each loss described in part \ref{chap:gscnn}. The sampling rates used in the ASPP module are 1, 6, 12, and 18 pixels. All those values were obtained by systematic testing. The best result, measured by Intersection Over Union (IOU, see \citealp{rezatofighi_generalized_2019}), was obtained after 675 iterations. 

\medskip

The first 195 models database was split for training and validation (training: 70\%, validation: 20\%, testing: 10\%). The second 150 models database was added to the training dataset to increase its variance. 

\medskip

Because GSCNN was developed to perform semantic segmentation of urban scenes from the Cityscape dataset \citep{cordts_cityscapes_2016}, some adjustments were made to apply the standard GSCNN architecture to the geological dataset. 

\medskip

The number of channels of the input image had to be reduced to 1 instead of 3. Moreover, the data augmentation module had to be adapted to the geological mapping task: the data were randomly rotated and resized, sheared, blurred, flipped vertically and horizontally to augment the variance of the training dataset in a geological-compatible way. The images are cropped to suppress the borders induced by shear and rotation.

\medskip

The validation set is also augmented by vertical and horizontal flip and 45 degrees rotations. Those transformations are combined for each validation data to have a high-variance case study and a consistent validation scenario. 

\medskip

Finally, a loss function of the standard GSCNN implementation, considering the quality of the segmentation along the contacts, was removed because of artifacts created by the data-augmentation module. 

\section{Results}

Figure \ref{fig:resu} shows the forward modeling response of the conceptual model (column A) and the response of three synthetic models obtained after data augmentation workflow (columns B, C, and D). Those models correspond to the simulations 1, 2, 3 shown in figures \ref{fig:MPS} and \ref{fig:SGS}. The original model was not used for training nor validation. 

\begin{figure}
\centering
\includegraphics[width=1\textwidth]{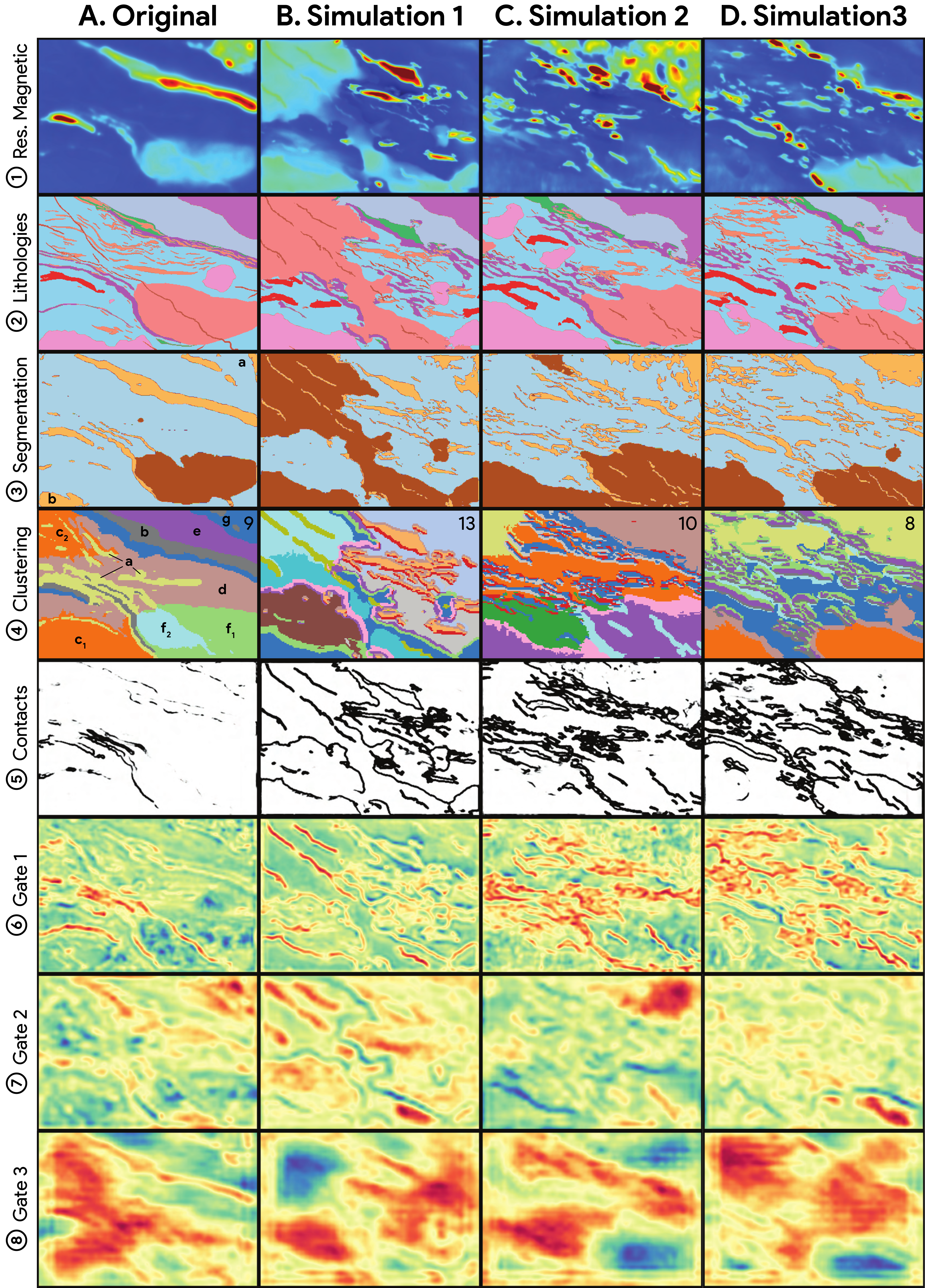}
\end{figure}
\begin{figure}
\centering
\caption{\label{fig:resu} Results of data augmentation and segmentation realized on Malartic model (A) and three synthesized models from figure \ref{fig:MPS} and \ref{fig:SGS} (B, C, D). The first row (1) presents the residual magnetic values used as inputs in the GSCNN. The second (2) row presents the geological mapping of the models, with 10 lithologies (the colors correspond to the \ref{fig:MPS}.1's legend). The third (3) row presents the segmentation output of the GSCNN with 3 lithological groups: the dyke like objects (yellow), the plutons (brown), and the greywackes (blue). The fourth (4) row presents the clustering result realized on the deep representations with the number of clusters used in the up-right corner. The fifth (5) row presents the results of the contact detection module, separating the input image in a binary image (contact/no contact). The sixth (6), seventh (7), and eighth (8) rows present the attention maps obtained from the GCL layers}
\end{figure}

\subsection{Data Augmentation Results}

The data augmentation workflow allows creating a large dataset of synthetic airborne magnetic data and associated lithological maps. 

\medskip

The first (1) row of figure \ref{fig:resu} shows the residual magnetic obtained by forward modeling on the Malartic magnetic susceptibility models (figure \ref{fig:resu}.1.A) and the first three synthetic magnetic susceptibility model (figure \ref{fig:resu}.1.A, \ref{fig:resu}.1.A, \ref{fig:resu}.1.A) presented in figure \ref{fig:SGS}. As shown in figure \ref{fig:resu}.1, the synthetic data (B, C, and D) are qualitatively comparable to real data (A). Plutons, dykes, ultramafic/volcanic rocks, and greywackes are well distinct. However, they show more variability than the original model, and the contacts between objects are sharper than the contacts of the original image. 

\medskip

The second (2) row of figure \ref{fig:resu} shows the surficial lithologies of the Malartic Model (figure \ref{fig:resu}.2.A) and the three synthetic lithological models (figure \ref{fig:resu}.2.B, \ref{fig:resu}.2.C, \ref{fig:resu}.2.D), presented in figure \ref{fig:MPS}. The contacts of these maps are used as contact masks. As figures \ref{fig:resu}.2.B, \ref{fig:resu}.2.C, and \ref{fig:resu}.2.D show, the geological bodies have shapes that could occur in a real geological setting. The lithologies respect the shapes and the textures of their \textquote{parent} lithologies, shown in figure \ref{fig:resu}.2.A.

\medskip

Because some geological objects are too small or because their magnetic susceptibility is similar to the surrounding rocks' signature, a part of the mapped lithologies are not visible on the residual magnetic results (see figure \ref{fig:original_model}.1 and \ref{fig:original_model}.2). In the original image (\ref{fig:resu}.2.A), those objects are the majority of diorite-monzodiorite dykes, granodiorite plutons, and some diabase dykes. 

\medskip

As described in part \ref{chap:gscnn} \textquote{Lithological Domain Simplification} some lithological domains had to be merged for the training process. The Pontiac's and Abitibi's greywackes were merged in a class, the granodiorite and diorite plutons in another one, and the diabase, mafic and ultramafic volcanics in a last one. Because iron formations are under-represented, they had to be merged with the ultramafic volcanics, as they present almost similar shapes and magnetic signatures in our model. At last, because the SGS module of the data augmentation mixed the distribution and variograms of the Diorite-Monzodiorite dykes and other dykes to avoid over-fitting, those lithologies are merged in the same class. Finally, only three classes remain for the training: the greywackes, the plutons, and the dykes. 

\subsection{GSCNN Results}

The GSCNN is trained to segement the 3 classes (greywackes, plutons, and dykes) using the synthetic data from the data augmentation workflow. It products the segmentation, contact detection, clustering, and attention maps described below.

\paragraph{Segmentation Results}

Figures \ref{fig:resu}.3 shows the segmentation output of a trained GSCNN on the corresponding inputs (figure \ref{fig:resu}.1). The blue corresponds to the greywackes class, the brown to the plutons class, and the yellow to the dykes class. The quantitative result for the segmentation validation is 0.687 IOU. It was calculated from synthetic models which were not used in the training process. The result on the original model presents a 0.543 IOU. 

\medskip

On the synthetic data (B, C, D in figure \ref{fig:resu}), the majority of visible objects in the input residual magnetic image are segmented. The non-visible objects are not detected. The Malartic formation, composed of ultramafic rocks (upper-right corner of each image, in figure \ref{fig:resu}.3), is not entirely detected. In the areas covered by this formation, the dyke class (yellow) is detected even if no particular anomalies are visible on the residual magnetic.

\medskip

The result obtained for the original model (figure \ref{fig:resu}.2.A) presents lower IOU and is qualitatively inferior. As figure \ref{fig:resu}.3.A shows, the diabase dyke crossing the pluton is not detected even if it is visible on residual magnetic data. The Malartic group (see figure \ref{fig:resu}.3.A.a) is not entirely detected and the southern granodiorite (\ref{fig:resu}.3.A.b) is classified as a dyke. 

\paragraph{Clustering Results}

Figures \ref{fig:resu}.4 show the results of the hierarchical clustering computed on the PCA calculated on the deep representation of the ASPP. The number of clusters selected for each model is indicated in the upper right corner of the images. It has been selected on a qualitative basis, where the balance between geological objects detection and segmentation clarity is considered the best. 

\medskip

The clustering result of the original image, shown in figure \ref{fig:resu}.4.A, is obtained by dividing the deep representations' pixels into 9 clusters. A part of the dykes was grouped in the same class (yellow, a in figure \ref{fig:resu}.3.A). Those dyke were detected from very subtle variations in the residual magnetic. They correspond to diorite-monzodiorite dykes and the northwest-southeast diabase dykes crossing the Pontiac greywackes. Most of the ultramafic and mafic formations are grouped in the same class (grey, b). The southern granodiorite is detected (orange, c\textsubscript{1}), and the Northwest part of the Pontiac greywackes (c\textsubscript{2}) is associated with this class. The rest of the Pontiac greywackes formation is detected as a unique class (light-brown, d). The Abitibi greywackes are also grouped in the same class (purple, e). The diorite is detected and split into two classes (light-blue and green, f\textsubscript{1} and f\textsubscript{2}). The delimitation between those two classes corresponds to the diabase dyke crossing the batholite. The last class (navy-blue, g) corresponds to high magnetic peak features, clustering the iron formation, the South of the CLLFZ, and the ultramafic Malartic group. 

\medskip

The results of the clustering for the simulations shown in figures \ref{fig:resu}.4.B, \ref{fig:resu}.4.C, and \ref{fig:resu}.4.D are obtained respectively with 13, 10, and 8 clusters. The plutons are globally well detected and grouped in distinct classes. Only the two northern granodiorites of simulation 2 are misclassified. The ultramafic, mafic, diorite-monzodiorite, and iron formations are grouped in the same classes, and the diabase dykes mostly in another one. The greywackes are grouped with the Malartic group in other classes.

\paragraph{Contact Detection}

The results of the contact detection are shown in figures \ref{fig:resu}.5. As shown in figures, the contact detection for the synthetic models is quite precise (\ref{fig:resu}.5.B, \ref{fig:resu}.5.C, and \ref{fig:resu}.5.D), almost as detailed as what a non-domain expert could accomplish. However, the contacts of the original images (\ref{fig:resu}.5.A) are not as well detected. 

\paragraph{Gates Representation}

Figures \ref{fig:resu}.6, \ref{fig:resu}.7, and \ref{fig:resu}.8 present the attention maps from the GCL layers. Those maps are 2D representations of the visual features associated with contacts, extracted from inputs data, at different scales. 

\medskip

The first GCL attention map (figure \ref{fig:resu}.6) isolates features associated with low-level contacts. It allows distinguishing very subtle variations in the residual magnetic. Those variations highlight most of the contact between different geological objects present in the geological map (figure \ref{fig:resu}.2), which were detected neither in the segmentation, the clustering, nor the contact detection module. 

\medskip

The second GCL attention map (figure \ref{fig:resu}.7) isolates medium-level features from the input image. Those features are mostly the contacts between the most significant geological objects (plutons) or volcanic formations. However, only a part of those types of contact is highlighted in this representation.

\medskip

The third GCL attention map (figure \ref{fig:resu}.8) isolates the high-level features from the input image. It separates the images into large areas. For example, it highlights the CLLFZ and separates the greywackes and the plutons. Those representations, however, do not precisely separate the input image into the known geology. 

\begin{figure}
\centering
\includegraphics[width=0.9\textwidth]{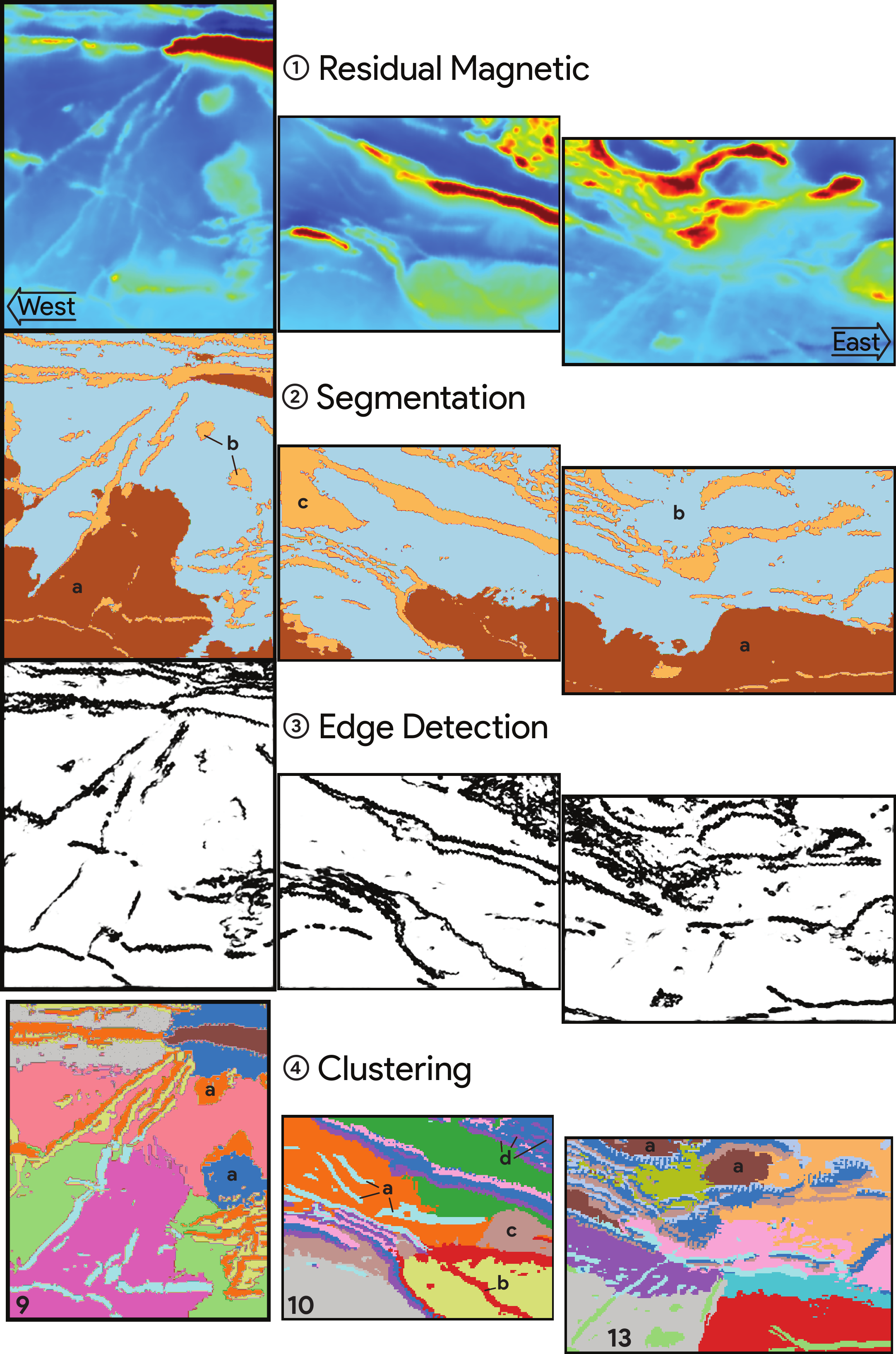}
\caption{\label{fig:tmr} 
Measured residual magnetic data of the areas at the West of Malartic (left), in Malartic area (center), and East of Malartic (right) (1); segmentation outputs obtains with the train GSCNN on those areas (2); contact detection outputs obtains with the shape stream in those areas (3); results of hierarchical clustering performed on the ASPP for those areas (4) with the number of cluster in the lower-left corner}
\end{figure}

\subsection{Transfer Learning: Testing The Network With Measured Data}

\begin{figure}
\centering
\includegraphics[width=0.9\textwidth]{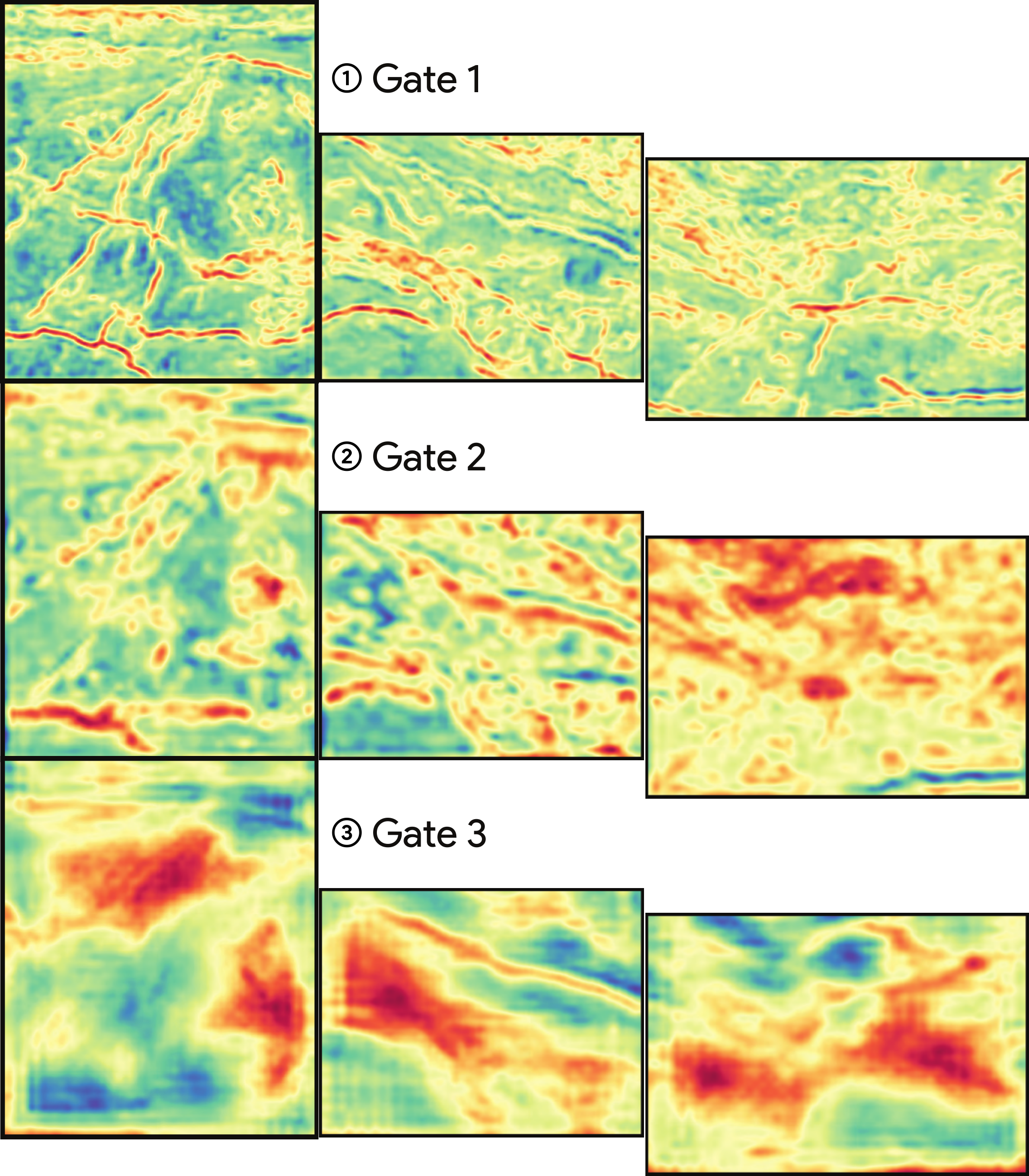}
\caption{\label{fig:tmg}
attention maps of the first (1), second (2) and third (3) GCL obtain with the measured residual magnetic data of the areas at the west of Malartic (left), in Malartic area (center), and east of Malartic (right)}
\end{figure}

The measured residual magnetic data of western and eastern extensions of the Malartic area are used as testing data in the transfer learning application. They correspond respectively to the area 1 and 3 in figure \ref{fig:localisation}, and are presented in figure \ref{fig:tmr}.1. Unlike synthetic models, only a qualitative approach is conducted because the ground truth is unknown.

\medskip

The western area is in the direct continuity of the Malartic Mine zone. It is mainly composed of greywackes belonging to the Pontiac formation, cut by late dykes and felsic intrusions (see geological map on figure \ref{fig:localisation}.1). The CLLFZ cut the area in the northern part. A highly magnetic zone is present on the eastern section of the fault (see figure \ref{fig:tmr}.1 West). To the North of the fault, some greywackes and volcanic rocks are observed. 

\medskip

The eastern area is also the continuity of the Malartic mine zone in Val d'Or area. The CLLFZ cuts the area in its middle, along a northwest-southeast direction (see geological map on figure \ref{fig:localisation}.3). In the North of the area, the Malartic group covers almost all the ground and presents high residual magnetic values (see figure \ref{fig:tmr}.1 East). Intermediate intrusions cuts the Malartic group, presenting low magnetic values. To the South of the CLLFZ, the Pontiac greywackes formation, cut by late dykes is observed.

\paragraph{Residual Magnetic}

The residual magnetic data were obtained from the \cite{sigeom_sigeom_2019} database. \cite{sigeom_sigeom_2019} is a compilation of the residual magnetic data at a 75 m resolution from the entire province of Québec. Some interpolation artifacts are visible in the data and produce local noise \citep{intissar_compilation_2014}. The noise is particularly present in the data from the East zone of Malartic around high magnetic anomalies (figure \ref{fig:tmr}.1 East). In order to have the measured data at the same resolution as the training dataset, measured magnetic maps were interpolated at a 50 m resolution with the bicubic algorithm. A Gaussian filter was then applied to the data to reduce noise. 

\medskip

The residual magnetic data used as testing inputs are shown in figure \ref{fig:tmr}.1. The images in the center correspond to the Malartic mine area. Those data are not the one simulated from the 3D model, as the one in figure \ref{fig:modelisation}, but are actual measured data. 

\paragraph{Segmentation}

The segmentation outputs of the testing data are shown in figure \ref{fig:tmr}.2. For the eastern and western areas, the GSCNN did not perform well on plutons. It classifies them as a part of Pontiac sediments (a in figure \ref{fig:tmr}.2), and some known plutons or plutons-like anomalies are classified as sediments or dykes (b). 

\medskip

The dykes are, in general, well detected, as most of the known dykes or dykes-like shapes. However, some areas are misclassified into dykes, like the northwest corner of the central area (figure \ref{fig:tmr}.2.c). In addition, some dykes shapes are not detected by the segmentation module. Therefore, the segmentation of the central area, at the Malartic model's location, is more accurate regarding the geological map of the area (figure \ref{fig:localisation}). 

\paragraph{Contact Detection}

Figure \ref{fig:tmr}.3 shows the output of the contact detection module. The module highlights the noise of the testing data. Still, most of the shapes are contoured, following the natural contact between objects. However, the most subtle contacts are still not detected. 

\paragraph{Clustering Results}

Figure \ref{fig:tmr}.4 shows the result of the clustering of the ASPP deep representations. Unlike the segmentation, the vast majority of objects are well detected. On the central image, the classification of the measured magnetic signal seems better than the classification of the magnetic data obtained by modeling. Almost every dykes (a in figure \ref{fig:tmr}.4 central), corresponding to subtle variations in the residual magnetic signal, are detected and classified. The dyke crossing the diorite intrusion (b) and the western pluton (c) between the CLLFZ and the diorite intrusion are detected. Ultramafic dykes in the Malartic Formation are also detected and segmented (d).

\medskip

In the western area, the clustering detects most of the objects visible in the residual magnetic image. Unlike in the segmentation output, the plutons (a in figure \ref{fig:tmr}.4 West) are detected, as most of the dykes. The sediments from Pontiac and Abitibi subprovinces are classified into different patches, even if no magnetic discontinuity is observed.

\medskip

The clustering of the eastern area detects most of the geological objects. The plutons intruding the Malartic formation (a in figure \ref{fig:tmr}.4 East) are classified, and the dykes are finely segmented. In the western area, the sediments are classified into different patches, without evidence of magnetic discontinuity. 

\paragraph{Gates Outputs}

figure \ref{fig:tmg} shows the outputs of the GCLs. The attention maps of the first GCL enhanced every subtle variation of the input data. As shown in figure \ref{fig:tmg}.1, contacts were detected, allowing us to visualize them well. 

\medskip

The attention maps from the second GCL are presented in figure \ref{fig:tmg}.2. As shown in the figure, those attention maps highlight medium wavelength contacts. However, the output of this GCL does not enhance every contact belonging to this wavelength. 

\medskip

The attention maps from the third GCL are presented in figure \ref{fig:tmg}.3. They highlight long-wavelength contacts. However, they do not seem to separate the area into known geology, except for the CLLFZ in the central and western zones and the eastern area's intermediate plutons. 

\section{Discussion}


Some areas of economic interest are highly explored, and 3D geological models are available at their location. The 3D modeling of an area's geology and petrophysical properties can be considered the highest detailed geological label. Even if some details diverge from reality (for example, the inversion smooths the data along the vertical axis), the amount of 3D data used for their creation allow them to be used as training data and labels with high confidence. However, because few 3D geological models exist, using them to train a CNN requires data augmentation. 

\medskip

Using MPS to augment the geological dataset shows promising results. The generated synthetic geological models are geologially realistic. However, some artifacts can be seen, forming stairs-like shapes at different places along lithological contacts. Furthermore, because MPS uses visual pattern recognition methods to generate the geological shapes, a redundancy with CNN might induce over-fitting. Also, the number of control points has to be high enough to avoid trivial geology but low enough to allow the algorithm to create models with significant variations between each other.

\medskip

The limitation to fill this synthetic geological model with magnetic susceptibility is that internal geological structures of geological bodies drive their magnetic response. Thus, SGS's results might not be realistic, as they do not follow such logic. Consequently, the results shown in figure \ref{fig:SGS} do not have the exact same aspect as the original magnetic susceptibility model and present sharper contacts and less realistic homogeneous magnetic susceptibility. 

\medskip

The airborne magnetic of the synthetic petrophysical model is computed by forward modeling. The process allows to smooth the data, and the results of the modeling look almost similar to measured magnetic data, as shown in figure \ref{fig:resu}.1. Because of the limited depth of the models, deep anomalies cannot be generated. As regional anomalies are removed from raw magnetic data to infer the residuals, the authors do not consider this as an issue. The performance of the network on measured magnetic data is a good indicator of the pertinence of the proposed data augmentation method. The computation time of the forward modeling is the main limitation of the proposed data augmentation method. 

\medskip

Once a large amount of synthetic airborne magnetic data and counterpart geological maps are generated, they can be used to train a CNN. We used GSCNN architecture as it presents multiple advantages. Indeed, because some lithologies had to be merged for data augmentation and training, a loss of information occurs. The Shape Stream compensates for this loss. Moreover, it compensates for the inconsistency due to the wide variety of shapes of geological objects. At last, the attention maps produced by the GCL highlighting textures at different scales offer pertinent representations of the input data, that can be used to interpret them. 

\medskip

The GSCNN's results for the synthetic validation data are quantitatively and qualitatively good for the segmentation, contact detection, and clustering module. However, the results obtained on the original model are not as good, especially for contact detection. It can be explained by the heterogeneous spatial distributions of the synthetic magnetic susceptibility models, induced by the SGS, which generate sharp contacts. However, on measured data, the contacts are well detected. It can also be explained by a difference in flight line altitude, making contact sharpers, or high-frequencies enhancement in measured data. 

\medskip

The segmentation performs well on the synthetic validation data. The dykes segmentation is visually good, probably because of their typical linear shapes. However, the Malartic group was partially detected (see figure \ref{fig:resu}.3.a), even if no magnetic anomalies can be seen. Moreover, a dyke is falsely detected instead of a pluton at the southwest corner of the original image (figure \ref{fig:resu}.3.b). Those errors might be caused by over-fitting, as the Malartic formation does not have a dyke shape. 

\medskip 

In the measured data, dykes are well detected (see figure \ref{fig:tmr}.2). However, greywackes and plutons are often misclassified (figure \ref{fig:tmr}.2.a), and some plutons or parts of the sediments classified as dykes (see figure \ref{fig:tmr}.2.b and \ref{fig:tmr}.2.c), especially in the western and eastern areas. The misclassification might be due to over-fitting. Only specific plutons and sediments had been sent to the GSCNN during its training, so it cannot detect this class in other areas. Those results highlight the limitations of transfer learning.

\medskip

The clustering module performs well on measured data. The classification of the sediments into different patches is due to their extensive representation thought the area. These results are qualitatively good and can be used as a preliminary geological map. Because of the pertinence of these results, the over-fitting described previously might be due to the lasts convolutional layers of the network. The shape stream might help the network to be resilient to over-fitting before the final segmentation. 

\medskip

The attention map of the first GCL layers is promising. It acts like both a high-pass filter and a first derivative, isolating all the subtle variations in the data. Thus, it can be used by geologists to detect small objects or contacts. However, the second and third attention maps from the other GCLs seem not as pertinent. It can be explained by the size of the training images, relatively small compared to the Cityscape dataset, and by the lack of medium and large wavelength anomalies in the synthetic airborne magnetic training data, due to the limited depth of the synthetic models. 

\section{Conclusion}

3D geological models are the highest level of knowledge of well-explored geological terrain. Therefore, it is pertinent to use them as a dataset to train a Convolutional Neural Network to do the preliminary mapping from airborne geophysical data. However, because of the restricted extent of those areas, a data augmentation method has to be implemented. 

\medskip

The proposed data augmentation workflow is a combination of three modules. Multi-Point Statistics creates synthetic geology, Sequential Gaussian Simulations populates the geological objects with petrophysical values, and forward modeling computes the airborne geophysical responses of the created models. Traditional data augmentation can then be used to augment the variance of the generated dataset. Because a Convolutional Neural Network performs well on these synthetic data, the method is empirically validated.

\medskip

Since predictive mapping is a segmentation task, Convolution Neural Network is the most pertinent algorithm to use, as they account for the spatial layout of the data. Furthermore, the GSCNN algorithm produces new pertinent representations of the input data. At last, the semi-supervised approach, using clustering on the convolutional neural network's in-depth features, shows quality results that can be used for preliminary geological mapping. Thus, it can be used in any area sharing geological structures with a 3D model and probably in foreign areas with similar geological context, using only airborne magnetic data.

\bibliographystyle{apalike}
\bibliography{biblio}
	
\end{document}